\documentclass[a4paper,11pt]{article}

\usepackage{amssymb}
\usepackage{amsmath}				
\usepackage{graphicx}
\usepackage{url}
\usepackage[numbers,sort&compress]{natbib}	
\usepackage[american]{babel}		

\begin{document}

\title{Counterflow Extension for the F.A.S.T.-Model}
\author{Tobias Kretz$^{1,3}$ and Maike Kaufman$^{2,3}$ and Michael Schreckenberg$^{3}$\\ \\
$^1$ PTV AG\\
Stumpfstra{\ss}e 1 -- D-76131 Karlsruhe -- Germany\\
E-mail: {\tt Tobias.Kretz@PTV.De}\\ \\
$^2$ Robotics Research Group \\ Department of Engineering Science \\ University of Oxford\\
Parks Road -- OX1 3PJ Oxford -- UK\\
E-mail: {\tt Maike@Robots.Ox.Ac.UK}\\ \\
$^3$ Physics of Transport and Traffic \\ University of Duisburg-Essen\\
Lotharstra{\ss}e 1 -- D-47057 Duisburg -- Germany\\
E-mail: {\tt Schreckenberg@PTT.Uni-DuE.De}\\
}

\maketitle

\begin{abstract}
The F.A.S.T. (Floor field and Agent based Simulation Tool) model is a microscopic model of pedestrian dynamics \cite{Schadschneider2009}, which is discrete in space and time \cite{Kretz2006f,Kretz2007a}. It was developed in a number of more or less consecutive steps from a simple CA model  \cite{Burstedde2001,Kirchner2002b,Kirchner2002a,Schadschneider2002a,Schadschneider2002b,Kirchner2003b,Nishinari2004}. This contribution is a summary of a study \cite{Kaufman2007} on an extension of the F.A.S.T-model for counterflow situations. The extensions will be explained and it will be shown that the extended F.A.S.T.-model is capable of handling various counterflow situations and to reproduce the well known lane formation effect.
\\
\noindent
\end{abstract}

\section{Introduction}
Counterflow situations \cite{Muramatsu1999,Helbing2000a,Muramatsu2000,Blue2000,Kirchner2002c,Tajima2002,Isobe2004,John2004,Nagai2005,Kretz2006a,Kretz2006d} pose a special problem to models of pedestrian dynamic as well as building construction. To avoid deadlocks a certain degree of self-organization is necessary, which typically leads to the formation of lanes, where people in the same lane follow each other and different lanes have opposing movement directions. CA models of pedestrian motion are particularly susceptible to deadlocks, if the main orientation axis of a corridor is aligned to one of the discretization axes. The reason for this is the given lane structure which leads to many head-on collision situation which lack any asymmetry to decide about the side of mutual passing. One modeling ansatz is to let opposing pedestrians exchange their cells with a certain probability \cite{Blue2001}. While this method is capable of reproducing some elements of reality, one still might be unhappy, as in reality people just do not walk through each other. A major temptation in modelling pedestrian counterflow using discrete models is to make use of the lanes preset by the CA lattice and align the two opposing groups' movement along one lattice axis. While such simulations can increase the understanding of a model or method, they are merely of academic interest and not sufficient for general use in safety, traffic or city engineering. The other big difficulty is combining counterflow situations with speeds larger than one cell per round.  The method presented here aims at being independent of the underlying lattice and it includes speeds larger 1.

\section{Extension of the F.A.S.T.-Model by a Comoving Dynamic Potential}
The probability $p_{xy}$ for an agent $i$ to select a cell $(x,y)$ as desired cell is multiplied by an additional factor $p^f_{xy}$:
\begin{eqnarray}
\bar{p}_{xy}&=&p_{xy}\cdot p_{xy}^f\\
p_{xy}^f &=& \exp{\left(k_f\sum_{j \in \mathcal{NN}_i}^N \frac{\sigma(i,j)}{|\sigma(i,j)|}P^j_{xy}\right)} \label{eq:1}
\end{eqnarray}
where $k_f$ is the coupling parameter that determines the strength of the effect. $\mathcal{NN}_i$ is the set of nearest neighbors of agent $i$. $N$ is the maximum number of nearest neighbors which is considered. $\mathcal{NN}_i$ only consists of agents within a distance $r_{max}$, which are visible to agent $i$ (i.e. not hidden by walls and in the field of view $\pi/2$ to the left and right of the direction of motion of agent $i$. $\sigma(i,j)=\vec{v}_i\cdot \vec{v}_j$ is the scalar product of the velocities $\vec{v}_i$ and $\vec{v}_j$ of agent $i$ and agent $j$, so the fraction in equation \ref{eq:1} is $+1$, if the agents rather move into the same direction and $-1$, if the rather move into opposite direction. Finally $P^j_{xy}$ is the value of the comoving potential induced by agent $j$ at position $(x,y)$. Assuming motion of agent $j$ at position $(x_0,y_0)$ along the x-axis, $P^j_{xy}$ has the following form (for any other direction of motion, the potential has to be rotated accordingly):
\begin{eqnarray}
P^j_{xy}&=&2h\left(\frac{v_j}{v_j^{max}}+\delta \right)\left(1-\frac{|y-y_0|}{a}\right) \text{, if } \frac{|y-y_0|}{|x-x_0|} \leq \frac{a}{b}\\ 
P^j_{xy}&=&2h\left(\frac{v_j}{v_j^{max}}+\delta \right)\left(1-\frac{|x-x_0|}{b}\right) \text{, if } \frac{|y-y_0|}{|x-x_0|} > \frac{a}{b}
\end{eqnarray}
where $h$ is the strength (height) of the potential. The content of the first brackets models the velocity dependence of the potential, $a$ is the basewidth of the potential orthogonal to the direction of motion, $b$ is the baselength alongside the direction of motion of agent $j$.

The values $N=12$, $h=4.0$, $\delta=0.2$, $a=2$, $b=r_{max}=15$, and $k_f=0.8$ have been established as a reasonable choice of parameters. Interestingly the simulation results worsened not only, if one reduced $N$, but also as it was increased. All simulations were done with $v_{max}=3$. Due to larger fluctuations the method works less well for speeds $v_{max}\geq 5$.

\section{Simulation Results}
\begin{figure}[htbp]
  \centering
	\includegraphics[width=0.8\textwidth]{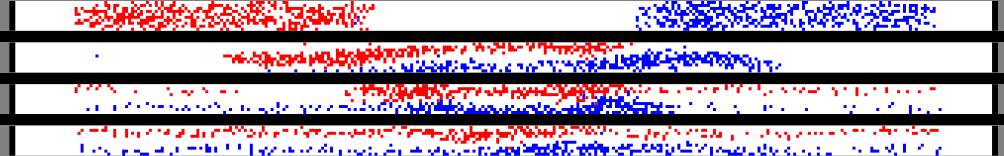} \vspace{12pt}\\
	\includegraphics[width=0.8\textwidth]{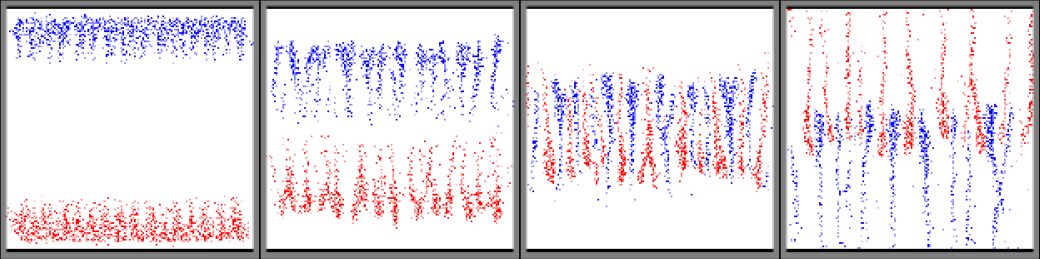} 
	\caption{Lane formation in a corridor and on an area.}
	\label{fig:results}
\end{figure}

\begin{figure}[htbp]
  \centering
	\includegraphics[width=0.39\textwidth]{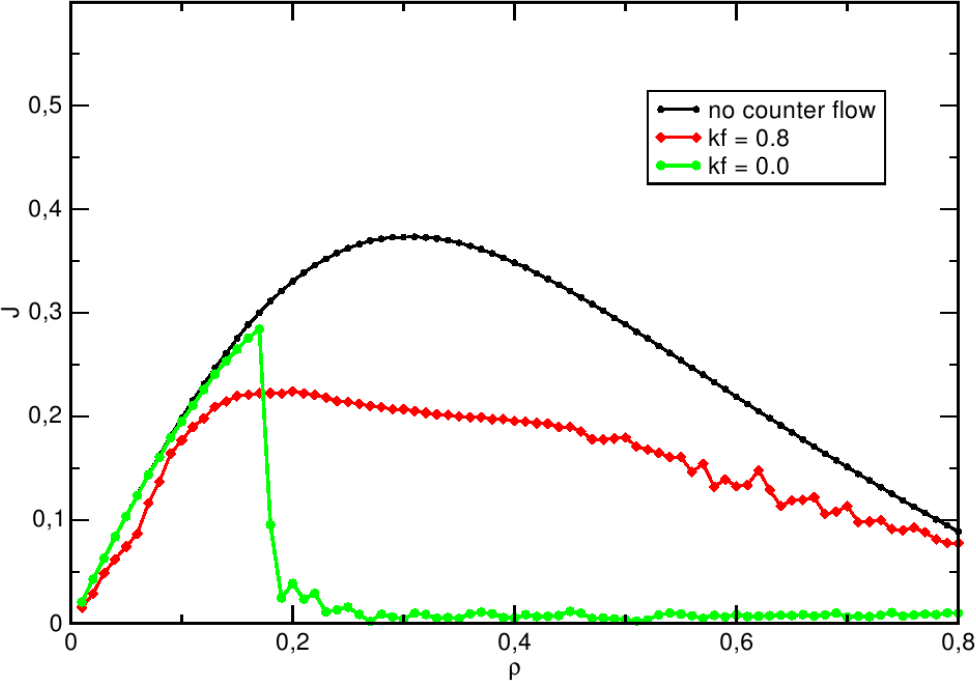} \hspace{12pt}
	\includegraphics[width=0.39\textwidth]{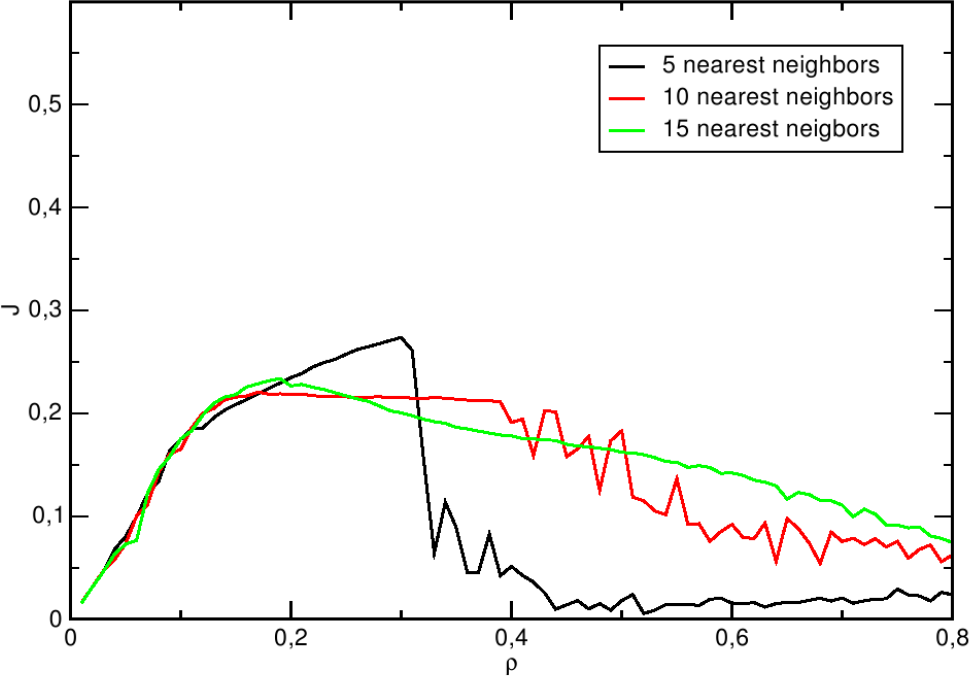} 
	\caption{Fundamental diagrams in dependence of $k_f$ and $\mathcal{NN}_i$.}
	\label{fig:results2}
\end{figure}

Figure \ref{fig:results} clearly shows, how lanes are formed and by that deadlocks are avoided and figure \ref{fig:results2} the effect on the fundamental diagram by variation of $k_f$ and $\mathcal{NN}_i$ is shown.

\section{Conclusions and Outlook}
This contribution presented a model extension to the F.A.S.T.-model for the simulation of counterflow. It was shown that lane formation was reproduced and deadlocks were avoided for speeds up to 3 cells per round and at densities well above the deadlock density of the earlier model. For even larger speeds the model needs to be stabilized, i.e. fluctuations must be restricted to some maximal value. An interesting empirical question would be to find the number of maximally considered nearest neighbors for real pedestrians.

\nocite{_PED2001,_TGF2001,_ACRI2002,_ACRI2006}
\bibliographystyle{unsrt}
\bibliography{Kretz-CCA-ACRI08}

\end{document}